\documentclass[aps,english,prb,reprint,superscriptaddress, citeautoscript,cite]{revtex4-1}

\usepackage{graphicx}
\usepackage{amssymb}
\usepackage{babel}
\usepackage{color}
\usepackage{multirow}

\begin{document}

\title{Effects of the number of layers on the vibrational, electronic and optical properties of alpha lead oxide.}
\author{Ali Bakhtatou}
\affiliation{Laboratoire de physique des mat\'{e}riaux, Universit\'{e} 8 Mai 1945 Guelma, BP 401 Guelma 24000, Algeria}

\author{Fatih Ersan}\email{fatih.ersan@adu.edu.tr}
\affiliation{Department of Physics, Adnan Menderes University, 09100,Ayd{\i}n, Turkey}
\affiliation{Department of Physics, University of Maryland Baltimore County, Baltimore, MD 21250, USA}

\begin{abstract}
We have investigated the effects of number of layers on the structural, vibrational, electronic and optical properties of $\alpha$-PbO using first principles 
calculations. Our theoretical calculations have shown that four Raman active modes of $\alpha$-PbO tend to red-shift from bulk to monolayer due to decreasing
of force constants and increasing of bond lengths. It has been shown that while bulk and multilayer $\alpha$-PbO have an indirect band gap, monolayer form
has a direct band gap value of 2.59 eV. Although lead atoms have 5d states, spin-orbit coupling does not significantly affect the band structure of 
$\alpha$-PbO. The computed cleavage energy value (0.67 J/m$^2$) confirms that monolayer PbO can be easily obtained from its bulk counterpart by exfoliation methods.
In addition to the band structure, we also calculated the optical properties and absorbed photon flux $J_{abs}$ of $\alpha$-PbO structures to investigate the 
possibility of solar absorption. Our calculations reveal that while monolayer and bilayer PbO have relatively larger band gaps and lower absorption 
coefficients, their $J_{abs}$ values are not ideal for solar absorption devices. In contrast, the multilayer and bulk phases of the $\alpha$-PbOs show good 
overlap with the solar spectrum and yield high electrical current values. Our calculations have indicated that ultrathin films of $\alpha$-PbO (such as 
3nm thickness) could be excellent candidates for solar cells. We believe that our work can be utilized to improve electronic and optical devices 
based on lead oxide structures. 


\end{abstract}

\maketitle
\section{Introduction} 
The synthesis of graphene~\cite{1}, the first two-dimensional material (2D), has attracted a great deal of attention because of its unique 
electronic, optical, and transport properties. It also has opened the path to experimental and theoretical research of other materials 
in monolayer or few monolayer form. A significant interest has been brought to nanosheets oxide (NS) 2D by the prediction of the first
metal oxide monolayer ZnO by Z. Tu \textit{et al.},~\cite{2} which was synthesized later by C. Tusche \textit{et al.}~\cite{3} Theoretical studies of this new 
material have revealed its interesting electronic and optical properties.~\cite{4,5,6,7,8} Subsequent studies have been carried out on monolayer 
oxide materials in the II-VI family, exhibiting promising dynamic stability of BeO, CaO, MgO, ZnO, CdO, and HgO semiconductors in addition 
to revealing attractive properties, which could be applied to nanoelectronics, optoelectronic and spintronics.~\cite{9,10} 2D MnO and NiO gain 
magnetization and exhibit different optical behavior from their bulk forms.~\cite{11} High electronegativity of the oxygen atom causes different
electronic structures for the monolayers of BO, AlO, GaO and InO, with respect to their S, Se and Te constituents.~\cite{12} Recently, a stable
direct band gap CO sheet with excellent mechanical properties was predicted to be a promising candidate for various applications.~\cite{13} 
Recently, $\alpha$-PbO 2D sheets have also attracted much attention. Previous studies have confirmed this oxide as one of the most useful materials
for optoelectronic sensing in the visible range.~\cite{14,15} Pasha \textit{et al.} concluded that stable nanostructures of lead oxides can also be 
obtained at low temperatures.~\cite{16} This material has an indirect band gap of about $\sim$1.9 eV in its bulk form.~\cite{17,18,19} Recent experimental
findings suggest that PbO atomic sheets exhibit hydrophobicity, thermal robustness, microwave stability, anti-corrosive behaviour and acid 
resistance.~\cite{20} Theoretical studies show that impurities of 3$d$ transition metal atoms in the $\alpha$-PbO (001) surface can induce magnetic moments 
to the structure.~\cite{21} Similar magnetization effects can be obtained by the nonmagnetic elements in group 13 and 14.~\cite{22} Possible point 
defects for $\alpha$-PbO has been studied by Berashevich \textit{et al.}, and their analysis shows that oxygen vacancies induce a deep donor level in the 
electronic band structure, while lead vacancies create a shallow level above the valence band. This implies that Pb vacancies act as an 
acceptor.~\cite{23} In this paper, we investigate the effects of dimensionality reduction from bulk to monolayer on its structural, vibrational,
electronic and optical properties. By means of density-functional theory using the generalized gradient approximation (GGA) of 
Perdew-Burke-Ernzerhof (PBE)~\cite{24} for exchange and correlation, we calculate the structural relaxed lattice parameters for the layered 
structures and compare their stability by analysis of the phonon spectrum of bulk, monolayer and bilayer. To overcome the problem of band gap 
underestimation of GGA, we employ Quasiparticle (QP) self-consistent GW (QSGW)~\cite{25} to compute and discuss the evolution of the electronic 
structure of monolayer, bilayer, trilayer, fourlayer and bulk $\alpha$-PbO and compare results with GGA. We find that the band gap in bulk and 
monolayer is significantly overestimated by this method; this is usually attributed to the underestimate of screening in the random phase
approximation (RPA). To remedy the overestimation, we adopt the hybrid method QSGW (hQSGW)~\cite{26} which has been shown to be an excellent 
predictor of electronic properties for a wide range of materials.~\cite{27} Finally, optical constants of monolayer, bilayer, trilayer, fourlayer
and bulk $\alpha$-PbO are computed and the absorbed solar photon flux $J_{abs}$ values (units of electrical current) are calculated.

\section {Computational Details}
To compute the lattice constants and the relaxed internal positions of the atoms for bulk and layered structures, we used the
projector augmented wave method (PAW) to deal with the electron-ion interaction and the GGA to approximate the exchange and correlation
part as implemented in VASP code.\cite{28} For electronic calculations, the mixed basis of plane waves and muffin-tin orbitals method 
(PMT)\cite{29} was used to solve the density functional Kohn- Sham equations in the GGA, the QSGW and the hQSGW approximations as implemented 
in the ecalj package.\cite{30} For bulk as well as for layered structures, the energy cutoff parameter for self-energy calculations is set 
to 3 Ry. For the augmented plane waves (APWs) of the PMT basis, the cutoff is set to 2 Ry. The basis set of the Muffin-tin orbitals (MTOs) 
is composed of the ($spdf$, $spd$) spherical harmonics of lead and of ($spd$, $sp$) spherical harmonics of oxygen. In addition, we treat Pb-$5d$ as 
local orbitals. Spin-orbit coupling is included in the end of the cycle iterations of QSGW and hQSGW. The k-point Brillouin-zone sampling 
(Monkhorst-Pack method)\cite{31} for the self-consistent calculations as well as for the QSGW self-energy calculations was performed 
using meshes of 6$\times$6$\times$4 for bulk, 6$\times$6$\times$2 for bilayer and 6$\times$6$\times$1 for monolayer. Since computing QSGW for a 
heavy elements like Pb (and for a 
number of atoms in the unit cell more than eight is very costly in time and memory), Brillouin zone meshing for GW was reduced to 3$\times$3$\times$1 k-points
for trilayer (12 atoms per unit cell) and 2$\times$2$\times$1 k-points for fourlayer (16 atoms per unit cell). For energy bands and partial density of 
states calculations, the number of k-points in the three reciprocal directions was chosen roughly in proportion to the size of the reciprocal 
lattice: 20$\times$20$\times$16 for bulk, 16$\times$16$\times$1 for monolayer, 12$\times$12$\times$2 for bilayer, 14$\times$14$\times$2 for trilayer and 
16$\times$16$\times$1 for fourlayer. To compute the 
frequency dependent dielectric functions in the independent particle picture we used the VASP\cite{28} code, where local field effects were 
not implemented in our calculations. For calculations of the optical properties, we used a $\Gamma$ centered 14$\times$14$\times$1 k-points mesh and 100 extra 
band numbers in addition to corresponding valence band numbers of each considered PbO structure. For the all VASP calculations, the kinetic
energy cutoff parameter was taken as 650 eV, the energy convergence value between two consecutive steps was chosen as 10$^{-5}$ eV, and a maximum
force on the each atom in the cell is allowed as 0.002 eV/\AA{} .

\section {RESULTS}
\subsection{Structural, Vibrational and Electronic Properties of $\alpha$-PbO}

\begin{figure*}
\centering
\includegraphics[scale=0.6]{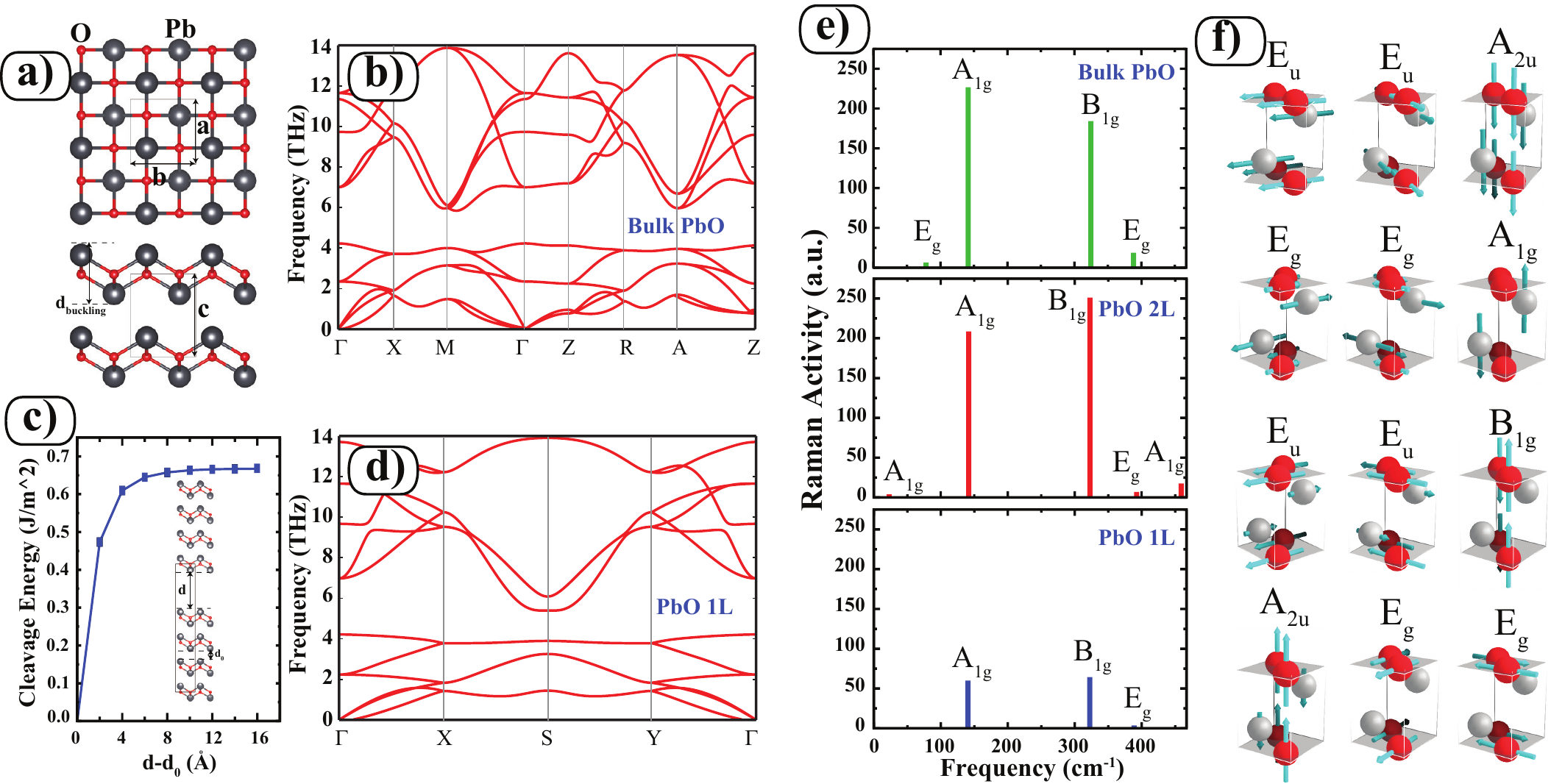}
\caption{(Color online) a) Top and side views of the optimized bulk $\alpha$-PbO structure. Primitive unit cell with the lattice constants a,b, and c is 
delineated by lines. Considered lentgh for the buckling paramater illustrated as $d_{buckling}$ in the structure. b) Calculated phonon dispersion curves for the 
bulk $\alpha$-PbO, c) Estimated cleavage energy curve, d) phonon dispersion of monolayer PbO, e) Computed Raman spectrums for the bulk, bilayer and monolayer 
forms of PbO structures and f) Vibratinol modes of Pb and O atoms at the center of the Brillouin Zone ($\Gamma$).}
\label{fig1}
\end{figure*}

We first carried out calculations on the bulk alpha lead oxide ($\alpha$-PbO) structure, which has P4/nmm space group. Fig.~\ref{fig1}a shows the $\alpha$-PbO structures.
The calculated equilibrium lattice constants are $a=b=4.047$ \AA\ and $c=5.121$ \AA\ . These values are only 2 $\%$ and 2.5 $\%$ larger than 
experimental results, for $a, b$ and $c$ respectively.\cite{32,33} The distance between the nearest Pb and O atoms is $d=2.35$ \AA\ and there is $2.72$ \AA\ between each 
successive PbO layer. For the sake of comparison, we calculated the phonon band structure of $\alpha$-PbO and the corresponding raman spectrum as seen in Fig.~\ref{fig1}. 
Our theoretical analysis indicates that the $\alpha$-PbO structure belongs to the D$^7_{4h}$ point group. There are 4 atoms per unit cell, therefore it has 3$\times$4=12 
phonon modes (Fig.~\ref{fig1}f), of which three acoustical (translational). The remaining 9 are optical (vibrational) modes. Irreducible representations of the $\alpha$-PbO 
at $\Gamma$, are as follows:

$\Gamma=4E_g+A_{1g}+2E_u+B_{1g}+A_{2u}$.

We see that seperately two phonon modes have the same vibration frequencies for the $E_g$ modes. Therefore, in fact $2E_g$ modes can be seen in the Raman spectrum for the 
$\alpha$-PbO structure. Among these irreducible representations, four are Raman-active modes ($2E_g$, $A_{1g}$ and $B_{1g}$), while the remaining 3 are 
infrared-active modes ($2E_u$ and $A_{2u}$). The theoretically calculated Raman spectrum of bulk PbO is illustrated in Fig.~\ref{fig1}e and all phonon modes are given in 
Table1. The most intense Raman mode at 141.24 cm$^{-1}$ ($A_{1g}$) corresponds to lead atoms moving in opposite directions parallel to the \textit{c}-axis. 
This calculated value is a bit smaller than the experimentally observed values of 145.5 cm$^{-}$ \cite{35} and 
150cm$^{-1}$ \cite{20}, or larger than 137 cm$^{-1}$ \cite{36}, but very close the result of 140 cm$^{-1}$ \cite{37}. These differences may be ascribable to 
temperature differences, since vibrational modes strongly depend on the materials temperature \cite{38}. The other strong intensity peak 
(324.47 cm$^{-1}$, $B_{1g}$) comes from the motion of oxygen atoms parallel to the $c$-axis.

One of the most favorable techniques to obtain few or monolayer structures from their layered bulk forms are liquid 
phase exfoliation or mechanical cleavage. Recently Kumar \textit{et al.} have produced few and single layer $\alpha$-PbO by using micromechanical, 
as well as sonochemical exfoliation, but this study holds does not provide the cleavage energy. Therefore we estimated cleavage energy of fourlayer PbO by 
creating a fracture in the eigth slab of $\alpha$-PbO structure and systematically increasing the distance between each successive fourlayer of PbO 
as seen in Fig.~\ref{fig1}c. The calculated cleavage energy is 0.67 J/m$^2$, and this estimated energy value is 1.7 times larger than the exfoliation energy of graphene 
(0.39 $\mp$ 0.02 J/m$^2$), but smaller than many estimated 2D materials such as GeP$_3$ \cite{39}, Ca$_2$N \cite{40}, NaSnP \cite{41}. Consequently, monolayers of 
$\alpha$-PbO can be produced by using exfoliation methods from its bulk form, as realized recently by Kumar \textit{et al.} \cite{20}. After these analyses, we created a 
monolayer PbO structure with the lattice constants of $a=b=4.022$ \AA\ and confirmed its dynamical stability by calculating the phonon frequencies along the all 
directions in its Brillouin Zone (Fig.~\ref{fig1}d). Phonon dispersion curves of monolayer PbO have only real modes over the whole BZ which indicate its stability. The highest optical frequency modes of monolayer 
PbO reach 13.91 Thz. This vibration mode value is very close to that of MoS$_2$ (14.18 THz),\cite{39} indicating the robustness of the covalent oxygen-oxygen bonds 
between the lead atoms. On the other hand, vibrational modes of lead atoms are under of 4.2 THz. We also checked thermal stability of the monolayer PbO by \textit{ab initio} 
molecular dynamic calculations for 2ps at 300K and 500K. Simulation results 
showed that the monolayer form of PbO can be stable at room and above temperature. In addition, we optimized the bilayer of PbO structure, 
which has an AA stacking order. The Raman-active modes and their corresponding intensities of monolayer and bilayer are compared with bulk PbO makes as illustrated 
in Fig.~\ref{fig1}e. The E$_g$ modes of bulk PbO at 78.50 cm$^{-1}$, which result from the motion of lead atoms parallel to the $xy$-plane, are redshifted to the 
76.52 cm$^{-1}$ for the bilayer with very weak intensity. Our results indicate this mode slips to 74.56 cm$^{-1}$ but its intensity goes nearly to zero,
which probably can not seen experimentally in Raman spectrum. In addition, the $B_{1g}$ mode tends to red-shift from bulk to monolayer PbO. The reason for 
these shifts can be explained by two types of interactions. First, the distance between the nearest Pb and O atoms for monolayer phase of PbO is $d=2.36$ \AA\, 
which is slightly larger than in bulk form. In principle, for the coupled oscilators, frequency ($\omega$) is related to the force constant ($k$) and the reduced 
mass ($\mu$) as $\omega=\sqrt{k/\mu}$. So, this increasing of the Pb-O bond distance appears in the Raman spectrum modes as a shift to lower frequencies due to 
decreased force constants of the Pb-O bonds. Second, $\alpha$-PbO has weak van der Waals interaction between 
the layers, and this vdW becomes weaker with decreasing number of layers, and disappears for the monolayer. For instance, while the perpendicular distance 
between the O-O atoms is 5.12 \AA{} for the bulk, this distance becomes 5.16 \AA{} for the bilayer PbO, so both long-range interaction and force 
constant decrease.

\begin{figure*}
\centering
\includegraphics[scale=0.66]{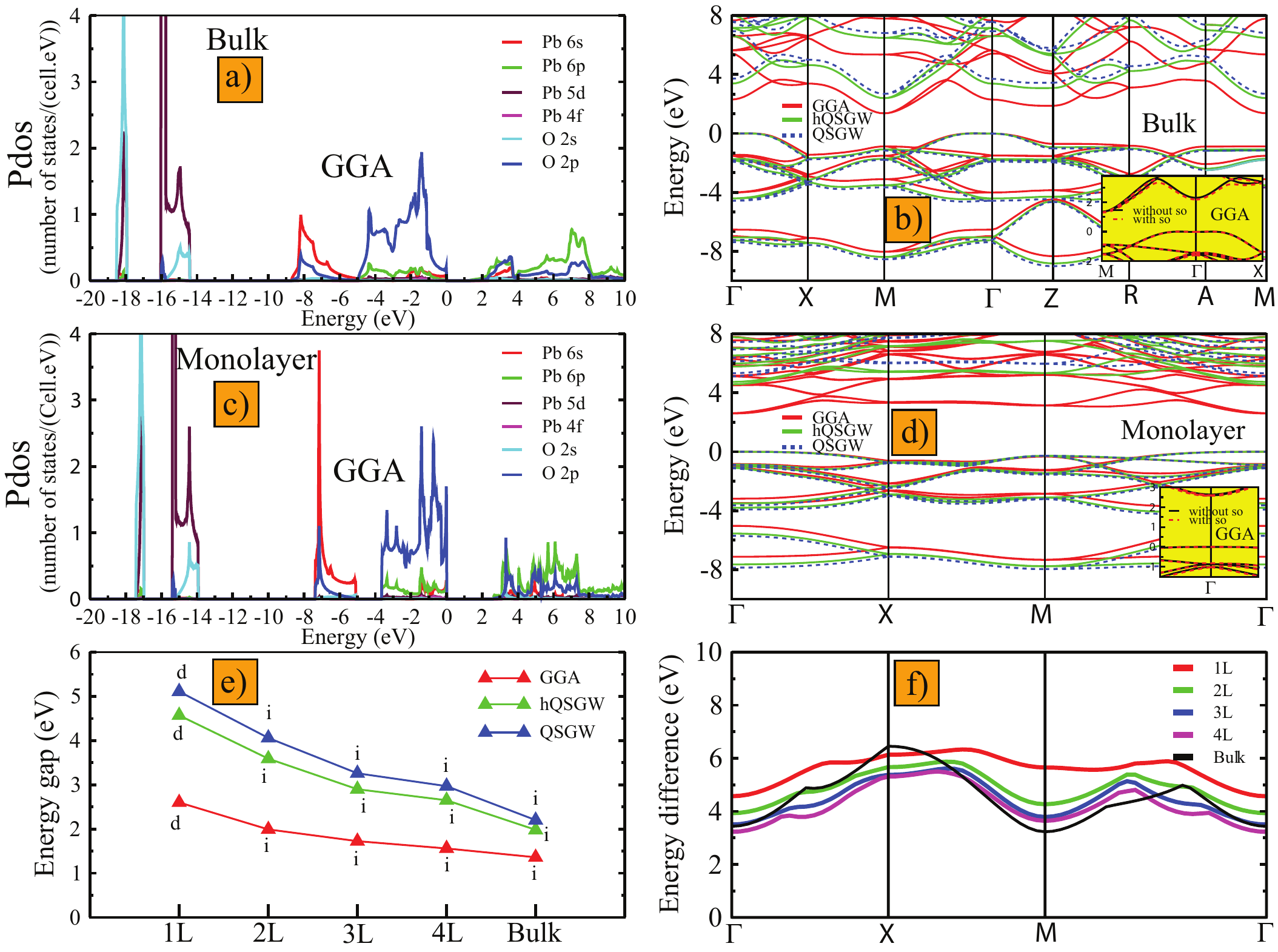}
\caption{(Color online) a), d) Electronic energy band structures and corresponding orbital projected densities of states of the bulk and monolayer PbO structures. 
e) Variation of the energy band gap with the number of layer of PbO. f) Variation of lower direct band gap value at every k-points through the $\Gamma$ to $\Gamma$.}
\label{fig2}
\end{figure*}

After the structural and stability analysis, we investigated the effects of the number of layers on the electronic properties of $\alpha$-PbO. 
For these examinations, we used two different density functional theory simulation packages (VASP and ecalj \cite{28,29}). Fig.~\ref{fig2}(a-d) 
presents the partial electronic density of states (PDOS) and band structures of bulk and layered $\alpha$-PbO structures, which are obtained from the ecalj 
package. Both VASP and 
ecalj packages give similar band structures and gap values for bulk or layered $\alpha$-PbO using the GGA method. Bulk $\alpha$-PbO has a 1.34 eV indirect 
band gap which increases to 2.20 eV by using the quasiparticle self-consistent GW (QSGW) method. We obtained good agreement with experiments 
(1.90, 1.95 eV) \cite{17,18,42} when we used hQSGW with spin orbit coupling (1.94 eV). Spin orbit coupling (SOC) only decreases the bulk $\alpha$-PbO 
band gaps by 40 meV (see Table 2 and Fig.~\ref{fig2} b). As can be seen from DOS of bulk $\alpha$-PbO in Fig.~\ref{fig2} a, oxygen $2p$ states dominate the valence 
band maximum but hybridize 
with lead $6s$ states, and the lowest energy regions are occupied by Pb $6s$ and O $2p$ orbitals. The main contribution to the conduction band minimum comes from the 
lead $p$ states. For the monolayer case of $\alpha$-PbO, the character of band stucture differs from the bulk. The band gap is 2.58 eV (for GGA results) 
and shows direct gap character as seen in Fig.~\ref{fig2} d. These GGA results are in good agreement with previous studies \cite{20}. Using QSGW, the band 
gap of the monolayer $\alpha$-PbO increases approximately two times from the GGA result (See in Table 2 and Fig.~\ref{fig2} d). The SOC effect is 
less important for the monolayer structure with respect to bulk $\alpha$-PbO; contribution of SOC only closes the band gap of monolayer PbO by 30 meV. 
The partial density of states also differs from bulk $\alpha$-PbO. The highest conduction bands are mainly Pb-$6p$ derived, and the contribution 
of O-$2p$ states is more important in monolayer than in bulk. The bands along the reciprocal directions at the edges of every subband in the valence band are 
flatter for monolayer than for bulk. Consequently the density of states at the edges of the four valence subbands are more significant than that of the bulk. 
For monolayer compared to bulk, the valence, the O-$2p$, the Pb-$6s$, the Pb-$5d$ and the O-$2s$ bandwidths are reduced by about 1.1 eV, 1.27 eV, 1.43 eV, 0.38 eV 
and 0.14 eV respectively. Owing to the flatter bands at the VBM and CBM, there are several pins in the DOS graphs. In addition, the Pb $6s$ and $6p$ states 
give nearly same contribution to the VBM, which is different from the bulk PDOS. We calculated effective mass values for holes and electrons at the 
VBM and CBM using interpolation techniques to obtain the slope of each curve. Our results show that holes are $\sim$ 9 times heavier than 
electrons for the bulk $\alpha$-PbO. For monolayer this ratio is $m^*_{h} \sim 63 m^*_{e}$. These values differ from a previous study, which found holes 
6 and 83 times heavier for bulk and monolayer PbO, respectively.\cite{43} This difference may come from the simulation packages used. As seen in Fig.~\ref{fig2} e, 
the band gap value does not increase linearly from bulk to monolayer, and only monolayer PbO has a direct band gap while all other 
layered PbO have indirect band gaps. Fig.~\ref{fig2} f shows the lower direct band gap value at every k points in the $\Gamma$-$X$, $X$-$M$ and $M$-$\Gamma$
reciprocal directions for monolayer, bilayer, trilayer, fourlayer and bulk $\alpha$-PbO. The monolayer direct band gaps are higher than those of bulk and 
other layered structures except around the $X$ point where the bulk direct band gap is the greatest . We can predict, from this figure, that the optical 
absorption onset of trilayer, fourlayer and bulk is approximatively the same and due to direct transitions at $M$ and $\Gamma$ k points, while the onset absorption of 
bilayer will be slightly above them (due to the same direct transitions), and the onset monolayer will be the highest and due only to direct $\Gamma$ transition.

\subsection{Optical Properties of $\alpha$-PbO}
\begin{figure*}
\centering
\includegraphics[scale=1]{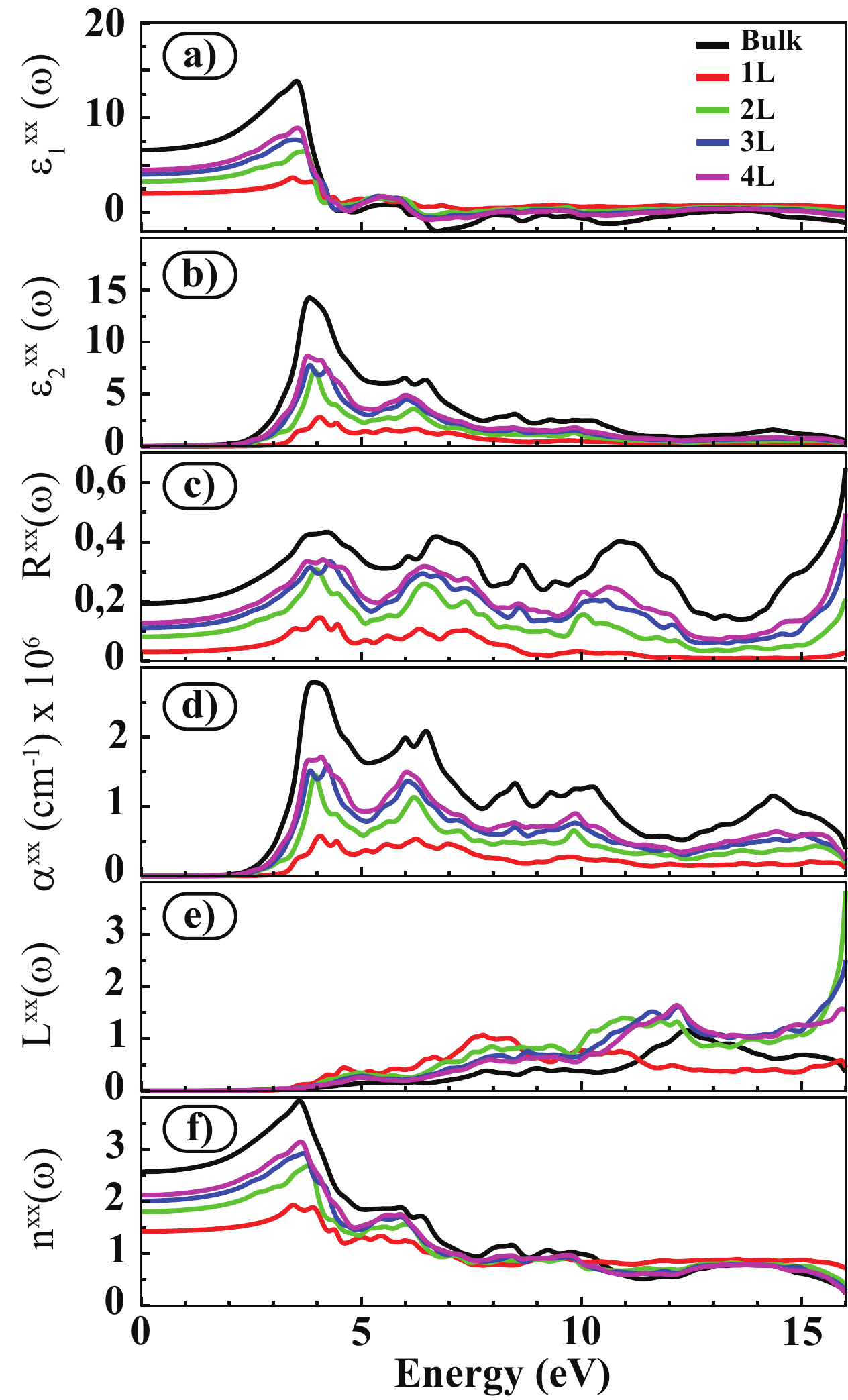}
\caption{(Color online) a) Real $\varepsilon_1(\omega)$ and b) imaginary $\varepsilon_2(\omega)$ parts of the dielectric response function, c) reflectivity $R(\omega)$, 
d) absorption coefficient $\alpha(\omega)$, e) energy-loss spectrum $L(\omega)$, and f) refractive index $n(\omega)$ of $\alpha$-PbO structures as a function of 
photon energy. All optical spectrum curves obtained for the $x$ component of the dielectric response functions.}
\label{fig3}
\end{figure*}

In this section we have investigated how the number of layers affects on the optical properties of the PbO material. Optical spectra of the materials can be 
revealed from the frequency dependent dielectric functions ($\varepsilon(\omega)=\varepsilon_1(\omega)+i\varepsilon_2(\omega)$). Detailed information can be found 
in literature \cite{44,45}. It should be noted that the excitonic effects are not included in this study. Fig.~\ref{fig3} illustrates the frequency dependent 
real ($\varepsilon_1(\omega)$) and imaginary ($\varepsilon_2(\omega)$) parts of the dielectric function and related optical spectral quantities for $\alpha$-PbO 
structures. Due to PbO structures having a square plane unitcell in the \textit{x}-\textit{y} axes, $x$ and $y$ components of the dielectric functions are equal 
for the same frequency, so we only plotted $x$ components of the dielectric functions. We can estimate the interior intra-optical excitations in the material 
by considering $\varepsilon_2(\omega)$ and the electronic DOS of the $\alpha$-PbO structures. As seen from $\varepsilon_2(\omega)$ spectra of the PbO 
structures, the threshold energies are comparable with the band gap values, which are given in Table 1. These threshold energy values are attributed to the 
interband transitions from O $2p$ states at the VBM to the lead $6p$ states at the CBM. Dominant peaks of $\varepsilon_2(\omega)$ around the 4 eV come from the 
excitations between the O $2p$ states in the valence bands to the O $2p$ or Pb $6p$ states in the conduction bands for all $\alpha$-PbO structures. As seen from 
the reflectivity spectra, PbO is a transparent material in the visible region, while reflection coefficients increase in the ultraviolet region. However, transparancy 
increases as layer decreases. As is known the peaks in the energy loss spectra can specify the collective excitations. 
Apparent peaks for the $\alpha$-PbO structures show up at around 12 eV, except for monolayer PbO where peaks exist at 7.5 eV. Refractive index of 
the materials decreases with the number of layers (see Fig.~\ref{fig3} f).

\begin{figure*}
\centering
\includegraphics[scale=0.7]{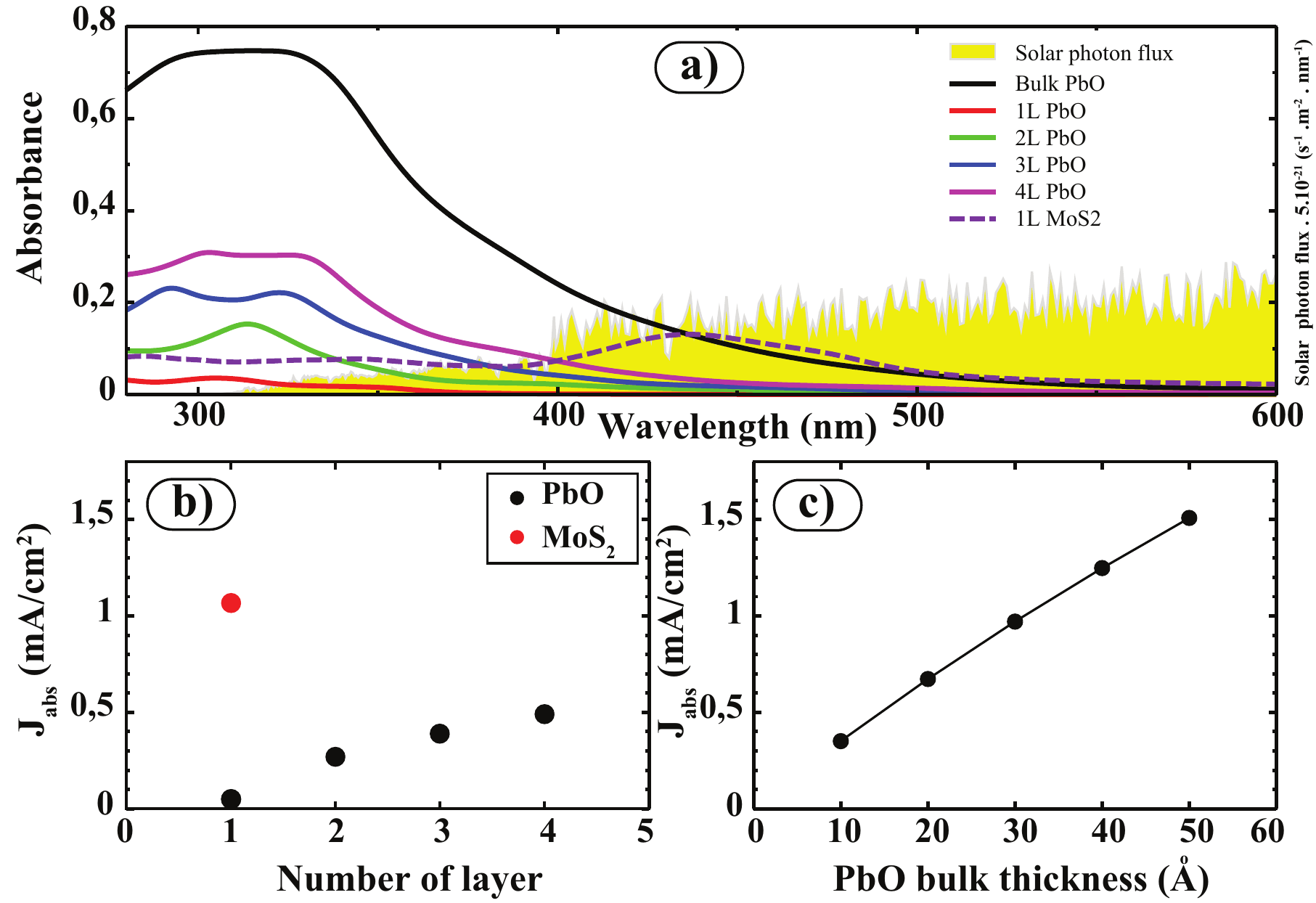}
\caption{(Color online) a) Absorbance (Derived from equation 1) of the $\alpha$-PbO and monolayer MoS$_2$ structures along $x$ direction. The photon flux of 
AM1.5G\cite{47} solar spectrum is shown in yellow. b) Variation of $J_{abs}$ as a function of the number of PbO layer. $J_{abs}$ value of monolayer MoS$_2$ 
is illustrated by red dot for the comparison. c) Computed $J_{abs}$ values for the bulk $\alpha$-PbO for various thicknesses. }
\label{fig4}
\end{figure*}

High absorption coefficients of the materials suggest that they can be used as highly efficient optical absorbers. As seen in Fig.~\ref{fig3} d, the $\alpha$-PbO 
structure has high absorption coefficients on the order of 10$^6$ cm$^{-1}$, which is 10 to 100 times larger than traditional solar cell materials \cite{46}. 
After calculation of absorption coefficients we calculated absorbance spectra for all $\alpha$-PbO (see Fig.~\ref{fig4} a). The photon flux of AM1.5G solar 
spectrum \cite{47} is shown in yellow. Monolayer MoS$_2$ can absorb up to 5-10$\%$ incident sunlight in a thickness of less than 1 nm and so could be utilized 
as a highly efficient solar absorber \cite{48}. Due to this knowledge we calculated the absorbance spectrum of monolayer MoS$_2$ to compare to PbO structures. 
Absorbance of the thin materials can be calculated as follows \cite{48,49};
\begin{equation}
 A(\omega)=\frac{\omega}{c}\varepsilon_2(\omega)\Delta z
\end{equation}
where, c is the speed of light, and $\Delta z$ is the length of the simulation cell in the layer-normal direction (thickness). This equation is valid if the 
material has the small thickness ($\Delta z \rightarrow 0$). This formula can be obtained after a Taylor expansion \cite{47,50} of the equation 
$A(\omega)=1 - e^{- \alpha . \Delta z}$.  
In this study we accepted the $\Delta z$ as the polarizable electronic thickness. For this we used d$_{buckling}$ of the PbO layer as illustrated in 
Fig.~\ref{fig1} a and the van der Waals atomic radius as below\cite{51};
\begin{equation}
 \Delta z= d_{buckling}(PbO) + 2r^{vdW}_{Pb}
\end{equation}
We only used the van der Waals atomic radius of lead atom instead of lead and oxygen atoms combined due to oxygen atoms sandwiched between the lead atoms.\cite{51} 
Fig.~\ref{fig4} a shows the absorbance of the $\alpha$-PbO structures and monolayer of MoS$_2$. As can be seen, visible region of the solar spectrum mostly 
overlaps with the absorbance curves of MoS$_2$ and bulk $\alpha$-PbO, as also indicated by their band gap values. Therefore they can absorb the incident sunlight. 
We remark that 2L, 3L and 4L $\alpha$-PbO structures have high absorbance value, their spectrums overlap with the low intensity flux region (ultraviolet 
region) of the solar spectra, due to their large band gap values. Figs.~\ref{fig4}b and ~\ref{fig4}c compare the absorbed solar photon flux $J_{abs}$ in 1L, 2L, 3L, 4L 
$\alpha$-PbO, monolayer MoS$_2$ and bulk $\alpha$-PbO films of various thicknesses. To gauge capacity of each structure as solar absorbers, we computed 
$J_{abs}$ by using the calculated absorbance using the integral \cite{52}; 
\begin{equation}
 J_{abs}= e \int_{0}^{\lambda(E_g)} A(\lambda) J_p(\lambda) d\lambda
\end{equation}
where the integration limit $\lambda(E_g)$ is the band gap of the corresponding structure, $J_p(\lambda)$ is taken from the AM1.5G solar 
spectrum \cite{47}. All $J_{abs}$ values obtained for the in-plane absorbance and absorption coefficients of the considered materials for the 
out-of-plane direction are lower than their in-plane values, so their $J_{abs}$ values will be small. As mentioned before, due to expensive 
computational times we did not consider excitonic effects in our optic calculations, which leads to an underestimated spectrum. Therefore, 
the computed $J_{abs}$ value for the 
monolayer MoS$_2$ (for the thickness of 6.5 \AA{}) is 1.1 mA/cm$^2$, which is smaller than calculated with using Bethe-Salpeter equation (BSE)
(3.9 mA/cm$^2$).\cite{48} But, our calculated $J_{abs}$ value for the monolayer 
MoS$_2$ is comparable with the results of Tan \textit{et al}. \cite{53}. They found the $J_{abs}$ is 2.3 mA/cm$^2$, because they took the 
$\Delta z = 18$ \AA{}. When we computed using the same $\Delta z$ value, we found the $J_{abs}$ equal to 2.5 mA/cm$^2$. Based on these considerations, 
we expect a larger $J_{abs}$ when using more sensitive for 2L, 3L, and 4L $\alpha$-PbO calculations such as BSE. On the other hand, bulk $J_{abs}$ value 
of the $\alpha$-PbO structure inreases with increasing film thickness, reaches and exceeds the 
$J_{abs}$ value of the MoS$_2$ monolayer after a thickness of 30 \AA{}. Especially, our calculations indicate that ultra-thin $\alpha$-PbO has high potential in 
solar cell applications.

\section{Conclusion}
In conclusion, our first-principles calculations show that single layer $\alpha$-PbO is a dynamically and thermally stable material. The
calculated cleavage energy value indicates that exfoliation methods can be suitable to obtain single layer $\alpha$-PbO. Raman active 
modes of $\alpha$-PbO tend to redshift with the decreasing of the number of layers. This is due to the increasing of the bond lengths and
decreasing of the force constants. In addition, one of the $E_g$ value of the $\alpha$-PbO modes which is corresponding to the motion of lead
atoms parallel to the $xy$-plane disappears by decreasing the layer number. While bulk and multilayer lead oxide structures have indirect
band gaps, monolayer PbO has direct gap and it has larger band gap value. Spin-orbit coupling is not so effective on the 
electronic band structures, contribution of SOC only decreases the band gaps of $\alpha$-PbOs by $\sim$40 meV. The high absorption 
coefficients of the $\alpha$-PbO structures make this material appealing for solar absorption applications. In spite of large band gaps of mono and 
bilayer PbO, other band gaps of multilayered
and bulk $\alpha$-PbO structures are in the visible region. Thereby, their obtained absorbance curves overlap with the solar photon flux spectra. 
The computed $J_{abs}$ value for single layer PbO is very small and not suitable for sunlight absorption. With the increasing of the number of 
lead oxide layers, the $J_{abs}$ value increases. With this increased $J_{abs}$ we believe that ultra-thin $\alpha$-PbO films can be excellent solar cell 
candidates. These theoretical results can be helpful in designing new solar absorption devices in future optoelectronic industry.

\section{Acknowledgments}

Computing resources used in this work were provided by the TUBITAK ULAKBIM, High Performance and Grid 
Computing Center (Tr-Grid e-Infrastructure) and by the "Plateau Technique de Calcul Intensif de Guelma".

\clearpage
\newpage
\begin{table*}
\centering
\caption{Raman-active (R.A) and Infrared-active (I.A) phonon frequencies in cm$^{-1}$: $\alpha$-PbO bulk, bilayer and monolayer. Comparison of the frequencies with 
the experimental, previously reported (Theoretical) and present study. }
\scalebox{0.67}{
\begin{tabular}{cccccc}
\hline
      & &  Bulk &                                           & Bilayer & Monolayer  \tabularnewline \hline
 Mode    &Present Study & Experimental & Theoretical.\cite{36} & Present Study & Present Study \tabularnewline \hline
E$_g$ (R.A)   & 78.41  &      &   & 76.51  & 74.50  \tabularnewline
E$_g$ (R.A)   & 78.50  & 81\cite{35}, 81\cite{34} & & 76.54  & 74.57  \tabularnewline
A$_{1g}$ (R.A)& 141.24 & 145.5\cite{35}, 140\cite{37}, 145\cite{34} & 142 & 22.91, 141.63, 458.94  & 140.49  \tabularnewline
E$_u$ (I.A)   & 233.58 &       &     & 234.51 & 232.75  \tabularnewline
E$_u$ (I.A)   & 233.59 &       &     & 234.54 & 232.76  \tabularnewline
B$_{1g}$ (R.A)& 324.47 & 337\cite{35}, 341\cite{37}, 340\cite{34} & 324 & 322.80 & 322.48  \tabularnewline
A$_{2u}$ (I.A)& 378.50 &         & 371 & 388.65 & 388.66  \tabularnewline
E$_g$ (R.A)   & 388.38 &         & 424 & 392.44 & 388.67  \tabularnewline
E$_g$ (R.A)   & 388.39 &         &     & 392.42 & 457.20  \tabularnewline  
\hline
\label{table1}
\end{tabular}
}
\end{table*}

\clearpage
\newpage
\begin{table*}
\centering
\caption{LDA, GGA, one-shot GW, hQSGW and QSGW gaps (with (w) and without(wo) spin-orbit-coupling (SOC) effects) and absorbed solar flux $J_{abs}$ 
(converted to units of equivalent electrical current (mA/cm$^2$)) of bulk, monolayer, bilayer, trilayer and fourlayer $\alpha$-PbO. The label 
one-shot means the 1shot GW calculation from the LDA result. The $J_{abs}$ for bulk in table is for a thickness of 50 \AA{}.}
\scalebox{0.58}{
\begin{tabular}{ccccccccccccc}
\hline
               & & & & & Gap Values (eV)& & &   & & &                         \tabularnewline \hline
               & LDA & & GGA & &one-shot GW & &hQSGW & &QSGW & &                         \tabularnewline \hline
               & wo-SOC & w-SOC & wo-SOC & w-SOC & wo-SOC& w-SOC& wo-SOC& w-SOC  &wo-SOC &w-SOC & $J_{abs}$ \tabularnewline \hline
 Bulk-PbO      &1.32 &1.27 &1.36 &1.31 &2.07 &2.03 &1.98 &1.94 &2.20 &2.16 & 1.51                         \tabularnewline
 1L-PbO        &2.55 &2.45 &2.59 &2.54 &4.79 &4.76 &4.57 &4.54 &5.11 &5.08 & 0.05                         \tabularnewline 
 2L-PbO        &1.94 &1.89 &1.99 &1.94 &3.78 &3.73 &3.59 &3.54 &4.06 &4.01 & 0.27                        \tabularnewline 
 3L-PbO        &1.68 &1.62 &1.72 &1.68 &2.99 &2.94 &2.90 &2.85 &3.26 &3.22 & 0.39                        \tabularnewline 
 4L-PbO        &1.52 &1.46 &1.56 &1.51 &2.77 &2.73 &2.65 &2.60 &2.98 &2.93 & 0.49                        \tabularnewline 
\hline
\label{table1}
\end{tabular}
}
\end{table*}

\clearpage
\newpage

\clearpage
\newpage

\clearpage
\newpage

\clearpage
\newpage

\end{document}